\documentclass[10pt, conference, compsocconf]{IEEEtran}

\usepackage{algorithm}
\usepackage{algorithmic}
\usepackage{tikz}
\usepackage{graphicx}
\usepackage{subcaption}
\usepackage{mathtools}
\usepackage{array}
\usepackage{cite}

\usetikzlibrary{automata, positioning, arrows}
\tikzset{->, >=stealth, node distance=3cm, every state/.style={thick, fill=gray!10}, initial text=$ $}
\let\OLDthebibliography\thebibliography
\renewcommand\thebibliography[1]{
  \OLDthebibliography{#1}
  \setlength{\parskip}{1.5pt}
  \setlength{\itemsep}{0pt plus 0.3ex}
}

\ifCLASSINFOpdf
\else
\fi
\begin{document}
%
\title{Energy-Efficient High-Throughput Data Transfers via Dynamic CPU Frequency\\ and Core Scaling}


\author{Luigi Di Tacchio\IEEEauthorrefmark{1}, Zulkar Nine\IEEEauthorrefmark{1}, Tevfik Kosar\IEEEauthorrefmark{1}, Fatih M. Bulut\IEEEauthorrefmark{2}, and Jinho Hwang\IEEEauthorrefmark{2}\\
\IEEEauthorrefmark{1}Department of Computer Science and Engineering, University at Buffalo, Buffalo, NY 14260\\
\IEEEauthorrefmark{2}IBM T.J. Watson Research Laboratory, Yorktown Heights, NY 10598}


%


\maketitle

\begin{abstract}
The energy footprint of global data movement has surpassed 100 terawatt hours, costing more than 20 billion US dollars to the world economy. Depending on the number of switches, routers, and hubs between the source and destination nodes, the networking infrastructure consumes 10\% - 75\% of the total energy during active data transfers, and the rest is consumed by the end systems. 
Even though there has been extensive research on reducing the power consumption at the networking infrastructure, the work focusing on saving energy at the end systems has been limited to the tuning of a few application level parameters such as parallelism, pipelining, and concurrency. In this paper, we introduce three novel application-level parameter tuning algorithms which employ dynamic CPU frequency and core scaling, combining heuristics and runtime measurements to achieve energy efficient data transfers. Experimental results show that our proposed algorithms outperform the state-of-the-art solutions, achieving up to 48\% reduced energy consumption and 80\% higher throughput.

\end{abstract}


%
\IEEEpeerreviewmaketitle

\section{Introduction}
\label{sec:introduction}
The tsunami of data generated by the Internet users, sensors, e-commerce, and surveillance cameras are fueling the large scale Artificial Intelligence (AI) systems. As a result, data transfer over Internet has been increasing exponentially each year and has already exceeded zettabyte scale~\cite{Cisco_2019}.
%
%
It is estimated that the communication industry could use 20\% of all world's electricity by 2025. More than one billion people are expected to come online in developing countries in the next 5 years. Currently, the data transfer task alone consumes hundred terawatt-hours energy with a price tag of 20 Billion US dollars annually. Moreover, the environmental side-effect is monumental - Information and Communication Technologies (ICT) alone could be responsible for a staggering 3.5\% carbon emission by 2020~\cite{theguardian}. This trend has motivated a considerable amount of work in energy consumption optimization of hardware and software systems as well as the network devices. 

Numerous works has been done to optimize the power consumption of the core network infrastructure (e.g., routers, switches, hubs, and network interface cards), however, little work has been done on energy efficiency in the end systems (i.e., sender/receiver compute nodes). Many of the works suggested to put idle components into sleep~\cite{Gupta_2003},turning off unused links, switches~\cite{heller2010elastictree}, link rate adaptation~\cite{lopez2014power}, power-aware optimization of packet routing based on power models~\cite{chabarek2008power}. 

Many of these approaches require expensive energy efficient hardware replacements. Some solutions require replacing existing protocol with a newer one, that comes with a huge adaptation overhead.  Most of the solutions cannot balance between performance and energy efficiency. Many of the solutions are not well accepted in the industry as they take a big toll out of the performance to provide energy efficiency. Moreover, these approaches hardly consider energy consumption in the end systems. In order to make the end systems more energy efficient, one can optimize the transport protocol with an adaptive sending rate. However, such a protocol update requires expensive kernel update. The adaptation of such protocol takes a long time as the operating system vendors show reluctance to such updates. In this paper, we propose a solution that is free from expensive hardware or protocol replacement, and provide a complete application layer solution which employs dynamic CPU frequency and core scaling. This novel solution can balance between performance and energy efficiency using Service Layer Agreement (SLA) based tuning. Our model is easy to deploy as it can be implemented in the user space. Moreover, user can set performance or energy constraints based on SLAs. 

The major contributions of this paper include the following:
\begin{enumerate}
    \item It proposes three novel application-level parameter tuning algorithms for energy efficient data transfers.
    \item It introduces CPU frequency scaling and active cores tuning to reduce the energy consumption of large scale data transfers.
    \item It combines heuristics and runtime measurements to dynamically and jointly tune five application-level parameters and satisfy the SLA requirements.
\end{enumerate}

The rest of the paper is organized as follows: Section~\ref{sec:parameters} provides a short description of the parameters that we aimed to optimize; Section~\ref{sec:techniques} provides an overview of the runtime parameter tuning model; Section~\ref{sec:algorithms} presents three SLA based energy-efficient parameter tuning algorithm; Section~\ref{sec:results} provides experimental evaluation of our models; Section \ref{sec:related_work} describes the related work in this field; and Section \ref{sec:conclusion} concludes the paper.

\section{Application-Level Parameters}
\label{sec:parameters}
Data transfer throughput and energy consumption are influenced by a plethora of parameters at different layers of the network protocol stack. However, there has been little work on tuning application layer parameters, which has the advantage of leaving the rest of the protocol stack unchanged.
To the best of our knowledge, this is the first work performing joint tuning of 5 application-level parameters: number of active cores, CPU frequency level, pipelining, parallelism, and concurrency.

\textit{Number of active cores} and \textit{CPU frequency level} determine the number of Instructions per Second (IPS) that the CPU can execute, as well as its energy consumption. Since pipelining, parallelism, and concurrency have an impact on the CPU utilization, it is important for those 3 parameters to be tuned jointly with the number of active cores and the CPU frequency level.

\textit{Pipelining} is the number of requests that can be sent back to back before having to stop and wait for the data to reach the destination. Pipelining reduces the total number of Round-Trip-Times (RTTs) required to complete the transfer: therefore, it is most beneficial when moving smaller files, since as the download or upload time decreases, the RTTs have a greater impact on the total transfer time.

\textit{Parallelism} is the number of file chunks that can be transferred concurrently for each file in the dataset. Using parallelism improves the network utilization by opening multiple connections and increasing the fraction of the bandwidth used during the transfer. Parallelism is most advantageous when transferring large files, especially when their size greatly exceeds the buffer size.

\textit{Concurrency} is the number of files that can be transferred at the same time on multiple connections. Like parallelism, opening multiple streams allows to use a larger share of the bandwidth, but it can improve the throughput even when transferring smaller files. Tuning concurrency is a difficult task, for having too many streams competing for a share of the bandwidth might lower the throughput and increase the energy consumption.

\section{Heuristic and Runtime Tuning Techniques}
\label{sec:techniques}
In this section, we present some of the techniques used to set and tune the 5 application-level parameters presented in section \ref{sec:parameters}. First, the transfer parameters are initialized using a heuristic-based approach, described in section \ref{subsec:heuristic}. After the transfer starts, runtime measurements are used to adjust the parameter values and guarantee the SLA requirements stipulated with the client, following the finite state machine presented in section \ref{subsec:fsa}. During the transfer, the CPU frequency and number of active cores are tuned following the threshold-based policy presented in section \ref{subsec:freqscaling}.

\label{subsec:heuristic}
\begin{algorithm}[ht]
    \caption{Heuristic-based parameter initialization}
    \label{alg:initialization}
    \begin{algorithmic}[1]
        \STATE{datasets = partitionFiles()}
        \FOR{dataset in datasets}
            \IF{avgFileSize $>$ BDP}
                \STATE{dataset.splitFiles(BDP)}
            \ENDIF
            \STATE{ppLevel = $\lceil$BDP / avgFileSize$\rceil$}
        \ENDFOR
        \STATE{tputChannel = avgWinSize / RTT}
        \STATE{numChannels = $\lceil$bandwidth / tputChannel$\rceil$}
        \FOR{dataset in datasets}
            \STATE{weight$_i$ = partitionSize$_i$ / $\sum_i$partitionSize$_i$}
            \STATE{ccLevel= $\lceil$weight$_i$ $\times$ numChannels$\rceil$}
        \ENDFOR
        \IF{SLApolicy(Energy)}
            \STATE{numActiveCores = 1}
            \STATE{coreFrequency = minFrequency}
        \ELSIF{SLApolicy(Throughput)}
            \STATE{numActiveCores = numCores}
            \STATE{coreFrequency = minFrequency}
        \ENDIF
    \end{algorithmic}
\end{algorithm}

\vspace{5mm}
\subsection{Heuristic-based Parameter Initialization}
Algorithm \ref{alg:initialization} is executed at the beginning of a data transfer. Its purpose is to initialize the 5 parameters described in section \ref{sec:parameters} to near-optimal values, based on a heuristic aimed at optimizing channels usage and utilization of system resources.

After clustering the data, if a partition contains files larger than the Bandwidth-Delay-Product (BDP), its files will be split in chunks (line 2-5). This has a twofold effect: multiple chunks can be transferred on different channels concurrently, increasing the throughput, and each chunk completely fills up the channel, its size being equal to the BDP.

Subsequently, the algorithm tries to find the minimum number of channels necessary to use the entire bandwidth (line 8-9). First, it calculates the theoretical throughput of a single TCP channel (line 8), as the average TCP window size over the Round-Trip-Time (RTT). The average window size can be easily estimated using a network benchmark tool such as iperf. Given the estimated channel throughput, the algorithm calculates the number of channels necessary to use the whole bandwidth (line 9).

After that, the channels are distributed among the file partitions based on the faction of data that each partition contains, with respect to the entire dataset (line 10-13).

Finally, depending on the SLA requirement, the algorithm initializes the number of active cores and the core frequency (line 14-20).


\subsection{Runtime Tuning Finite State Machine}
\label{subsec:fsa}
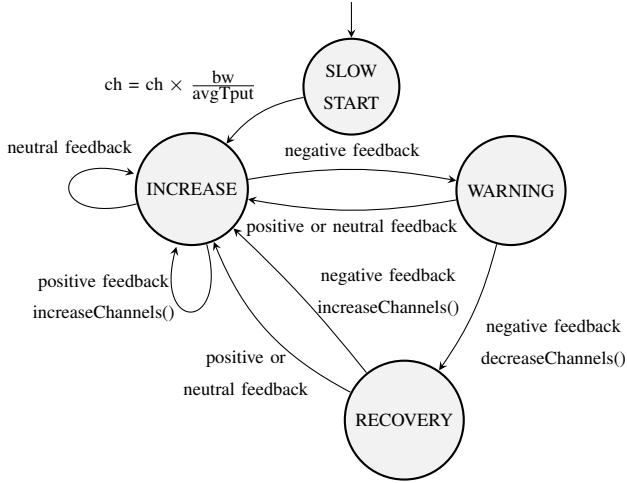
\begin{figure}[ht]
    \centering 
    \begin{tikzpicture}
    
        \node[initial above, state, anchor=north, align=center] (slow_start) {\scriptsize SLOW \\ \scriptsize START};
        
        \node[state, below left of=slow_start, anchor=south, align=center] (tuning) {\scriptsize INCREASE};
        
        \node[state, below right of=slow_start, anchor=south, align=center] (warning) {\scriptsize WARNING};
        
        \node[state, below right of=tuning, anchor=south, align=center, yshift=-5.0em, xshift=+2.0em] (recovery) {\scriptsize RECOVERY};
        
        \draw (slow_start) edge[bend right=20] node[anchor=west,above,xshift=-3.0em]{\scriptsize $\text{ch = ch} \times \frac{\text{\scriptsize bw}}{\text{\scriptsize avgTput}}$} (tuning)
        
              (tuning) edge[loop left] node[anchor=south,above,yshift=+1.0em]{\scriptsize neutral feedback} (tuning)
              (tuning) edge[loop below] node[anchor=left, left, align=center,xshift=-0.3em,yshift=+0.5em]{\scriptsize positive feedback \\ \scriptsize increaseChannels()} (tuning)
        
              (tuning) edge[above, bend left=10] node[anchor=west,above,align=center]{\scriptsize negative feedback} (warning)
              (warning) edge[below, bend left=10] node[anchor=east, below, align=center]{\scriptsize positive or neutral feedback} (tuning)
              
              (warning) edge[bend left=10] node[anchor=east, below, align=center, xshift=+3.0em]{\scriptsize negative feedback \\ \scriptsize decreaseChannels()} (recovery)
              
              (recovery) edge[bend right=4] node[anchor=west,align=center,xshift=+0.2em,yshift=+0.2em]{\scriptsize negative feedback \\ \scriptsize increaseChannels()} (tuning)
              (recovery) edge[bend left=20] node[anchor=east, below, align=center,xshift=-0.8em,yshift=-0.4em]{\scriptsize positive or \\ \scriptsize neutral feedback} (tuning)
        ;
    \end{tikzpicture}
    \caption{Algorithms Finite State Machine}
    \label{fig:fsm}
\end{figure}

\begin{algorithm}[t]
    \caption{Slow Start algorithm}
    \label{alg:slowstart}
    \begin{algorithmic}[1]
        \FOR{Timeout}
        	\STATE{calculateThroughput()}
            \STATE{numCh $*$= bandwidth / lastThroughput}
            \STATE{updateWeights()}
            \FOR{dataset in datasets}
            	\STATE{ccLevel$_i$ = weight$_i$ $*$ numCh}
				\STATE{updateChannels()}
            \ENDFOR
        \ENDFOR
    \end{algorithmic}
\end{algorithm}

After initializing pipelining, parallelism, concurrency, CPU frequency, and number of active CPU cores using algorithm \ref{alg:initialization}, the data transfer is started and a different algorithm is executed depending on the SLA agreement. Nonetheless, all three algorithms presented in this paper follow a similar structure that can be described using a Finite State Machine, illustrated in figure \ref{fig:fsm}.

The first state, denominated \textit{Slow Start}, is entered after algorithm \ref{alg:initialization}.
After a short timeout, the tuning algorithm measures the throughput and, if necessary, adjusts the number of channels to compensate for the initial estimation error.

In state \textit{Increase}, the transfer parameters are increased or decreased based on the feedback from the channel. If the algorithm's goal is energy-related, the feedback is represented by the energy consumption since the last timeout, otherwise it is the average throughput during the last time interval.

Upon receiving negative feedback, the algorithm transitions to the state \textit{Warning}. From there, a positive or neutral feedback suggests that the performance drop was only temporary, which causes a transition back to state \textit{Tuning}. However, upon receiving a second negative feedback, the algorithm enters state \textit{Recovery}.

From here, a positive or neutral feedback is a sign that reducing the channel count eased the load on the network, and the algorithm goes back to state \textit{Increase}. On the other hand, a negative feedback indicates that the channel's available bandwidth dropped, hence the previous channel count is restored and the algorithm transitions back to state \textit{Increase}.


\subsection{Threshold-based dynamic frequency and core scaling}
\label{subsec:freqscaling}
\begin{algorithm}[ht]
    \caption{Load Control algorithm}
    \label{alg:loadcontrol}
    \begin{algorithmic}[1]
        \FOR{Timeout}
        	\IF{cpuLoad \textgreater\ maxLoad}
        	    \IF{numActiveCores \textless\ numCores}
        	        \STATE{increaseActiveCores()}
        	    \ELSIF{cpuFreq \textless\ maxFreq}
        	        \STATE{increaseFrequency()}
        	    \ENDIF
        	\ELSIF{cpuLoad \textless\ minLoad}
        	    \IF{cpuFreq \textgreater\ minFreq}
        	        \STATE{decreaseFrequency()}
        	    \ELSIF{numActiveCores \textgreater\ 1}
        	        \STATE{decreaseActiveCores()}
        	    \ENDIF
            \ENDIF
        \ENDFOR
    \end{algorithmic}
\end{algorithm}

The CPU frequency and the number of active cores are dynamically tuned using a threshold-based policy, implemented in algorithm \ref{alg:loadcontrol}.

When the CPU utilization increases above a certain threshold, named \textit{maxLoad}, the algorithm tries to increase the number of active cores or CPU frequency, in order to reduce the load on the system (line 2-7). Conversely, if the CPU utilization is lower than a certain threshold, named \textit{minLoad}, the algorithm tries to reduce the CPU frequency or the number of active cores.

Algorithm \ref{alg:loadcontrol} is called at regular intervals by the parameter tuning algorithms to keep the energy consumption as low as possible without sacrificing performance. In fact, every time one of the other transfer parameters is modified, the CPU load might change as well, and could either use more energy than needed or cause a lower performance gain than expected. 

\section{Parameter Tuning Algorithms}
\label{sec:algorithms}
In this section, we present three novel energy-efficient parameter tuning algorithms, which dynamically adapt the parameter values to achieve three different SLA requirements: minimum energy consumption, maximum throughput, and target throughput.

\subsection{Minimum energy algorithm}
\label{subsec:minenergy}

\begin{algorithm}[h]
    \caption{Minimum energy algorithm}
    \label{alg:minenergy}
    \begin{algorithmic}[1]
        \STATE{SlowStart()}
        \FOR{Timeout}
        	\STATE{calculateThroughput()}
			\STATE{calculateEnergy()}
			\STATE{remainTime = remainData / avgThroughput}
			\STATE{predictedEnergy = avgPower $\times$ remainTime}
			\IF{state = INCREASE}
				\IF{E$_{last}$ + E$_{future}$ $<$ (1-$\alpha$) $*$ E$_{past}$}
					\STATE{numCh = min(numCh + $\Delta$Ch, maxCh)}
				\ELSIF{E$_{last}$ + E$_{future}$ $>$ (1+$\beta$) $*$ E$_{past}$}
					\STATE{state = WARNING}
				\ENDIF
				
			\ELSIF{state = WARNING}
				\IF{E$_{last}$ + E$_{future}$ $<=$ (1+$\beta$) $*$ E$_{past}$}
					\STATE{state = INCREASE}
				\ELSE{}
					\STATE{numCh = max(numCh - $\Delta$Ch, 1)}
					\STATE{state = RECOVERY}
				\ENDIF
			\ELSIF{state = RECOVERY}
				\IF{E$_{last}$ + E$_{future}$ $<=$ (1+$\beta$) $*$ E$_{past}$}
					\STATE{state = INCREASE}
				\ELSE{}
					\STATE{numCh = min(numCh + $\Delta$Ch, maxCh)}
					\STATE{state = INCREASE}
				\ENDIF
			\ENDIF
			
			\STATE{updateWeights()}
            \FOR{dataset in datasets}
            	\STATE{ccLevel$_i$ = weight$_i$ $*$ numCh}
				\STATE{updateChannels()}
            \ENDFOR
        \ENDFOR
    \end{algorithmic}
\end{algorithm}

The minimum energy algorithm tries to achieve minimum energy consumption using two different strategies: 1) increasing the concurrency level only if that results in a lower estimated energy usage; 2) increasing the active core count and the CPU frequency only if the CPU is reaching full utilization, while reducing the number of active cores and CPU frequency if the CPU load is lower than a certain threshold.

During the \textit{Slow Start} phase (line 1-9), the algorithm updates the weights assigned to each datasets, based on the remaining data size, and redistributes the channels across the datasets (line 7-10)

Once every timeout, the algorithm assesses whether or not the last parameter change has caused a performance improvement (line 10-14). Depending on the feedback, it either increases the channel count (line 17) or enters state \textit{Warning} (line 19).

While in state \textit{Warning}, the algorithm tries to assess whether the performance drop has been caused by an excessively high concurrency level or by a change in available bandwidth (line 21-27). In order to do that, it estimates the transfer energy consumption, and if still higher than the previous estimate, it decreases the channel count and moves to state \textit{Recovery}. If the energy spike was only temporary (line 22), it goes back to state \textit{Increase}.

In state \textit{Recovery}, the algorithm determines whether or not the parameter reduction has caused the energy consumption to lower (line 29). If that is the case, the previous channel count was too high and the new value is closer to optimal than the previous one. Otherwise (line 32), the available bandwidth has changed and the algorithm shifts back to state \textit{Increase} to find a new optimal channel count.

At every iteration, the algorithm recalculates weights and concurrency levels for each dataset based on the remaining data size. Slower datasets will receive a higher fraction of channels in order to complete the transfer at approximately the same time.


\subsection{Energy-efficient maximum throughput algorithm}
\label{subsec:maxtput}

\begin{algorithm}[h]
    \caption{Energy-efficient maximum throughput algorithm}
    \label{alg:maxtput}
    \begin{algorithmic}[1]
        \STATE{SlowStart()}
        \FOR{Timeout}
        	\STATE{calculateThroughput()}
			\IF{state = INCREASE}
				\IF{avgTput $>$ (1+$\beta$) $*$ refTput}
					\STATE{numCh = min(numCh + $\Delta$Ch, maxCh)}
					\STATE{refTput = avgTput}
				\ELSIF{avgTput $<$ (1-$\alpha$) $*$ refTput}
					\STATE{state = WARNING}
				\ENDIF
				
			\ELSIF{state = WARNING}
				\IF{avgTput $>=$ (1-$\alpha$) $*$ refTput}
					\STATE{state = INCREASE}
				\ELSE{}
					\STATE{numCh = max(numCh - $\Delta$Ch, 1)}
					\STATE{state = RECOVERY}
				\ENDIF
			\ELSIF{state = RECOVERY}
				\IF{avgTput $>=$ (1-$\alpha$) $*$ refTput}
					\STATE{state = INCREASE}
				\ELSE{}
					\STATE{numCh = min(numCh + $\Delta$Ch, maxCh)}
					\STATE{state = INCREASE}
					\STATE{refTput = avgTput}
				\ENDIF
			\ENDIF
			
			\STATE{updateWeights()}
            \FOR{dataset in datasets}
            	\STATE{ccLevel$_i$ = weight$_i$ $*$ numCh}
				\STATE{updateChannels()}
            \ENDFOR
        \ENDFOR
    \end{algorithmic}
\end{algorithm}

The energy-efficient maximum throughput algorithm tries to maximize the throughput while keeping the number of channels as low as possible. It reaches this goals by avoiding to increase the channel count if doing so does not increase the throughput.

The algorithm starts by executing the \textit{Slow Start} phase (line 1-10). It also updates the reference throughput to the average throughput measured in the \textit{Slow Start} phase. The reference throughput is set to the best achieved throughput in state \textit{Increase}, and is used to determine the feedback received from the channel while in other states.

Initially, the algorithm starts in state \textit{Increase}. Upon timeout, it measures the average throughput, and if higher than the reference throughput by at least a factor $\beta$, it increases the number of channels and updates the reference throughput (line 11-16). Otherwise, if the feedback was negative, it enters state \textit{Warning} to determine whether the performance drop has been caused by an excessively high channel count or by a change in available bandwidth (line 17-19).

If the algorithm receives positive or neutral feedback while in state \textit{Warning}, it goes back to state \textit{Increase}, assuming the performance drop was only temporary (line 21-23). Conversely, a negative feedback causes the algorithm to enter state \textit{Recovery} and temporarily reduce the number of channels (line 24-26).

If decreasing the channel count improved the throughput, the algorithm transitions to state \textit{Increase} without further changing the parameter values (line 28-30); otherwise, it restores the previous channel count, assuming that the performance drop was caused by a reduction in available bandwidth, and updates the reference throughput to the last measured average throughput (line 31-34).

Finally, no matter which state the algorithm is in, the weights for each datasets are recalculated based on the remaining data size, and channels are redistributed among datasets based on their weights (line 37-39).


\subsection{Energy-efficient target throughput algorithm}
\label{subsec:targettput}

\begin{algorithm}[h]
    \caption{Energy-efficient target throughput algorithm}
    \label{alg:targettput}
    \begin{algorithmic}[1]
        \STATE{SlowStart()}
        \FOR{Timeout}
        	\STATE{calculateThroughput()}

			\IF{state = INCREASE}
				\IF{avgTput $>$ (1+$\beta$) $*$ targetTput \textbf{or} \\
				 avgTput $<$ (1-$\alpha$) $*$ targetTput}
					\STATE{state = RECOVERY}
				\ENDIF
				
			\ELSIF{state = RECOVERY}
				\IF{avgTput $>$ (1+$\beta$) $*$ targetTput}
					\STATE{numCh = max(numCh - $\Delta$Ch, 1)}
				\ELSIF{avgTput $<$ (1-$\alpha$) $*$ targetTput}
					\STATE{numCh = min(numCh + $\Delta$Ch, maxCh)}
				\ENDIF
				\STATE{state = INCREASE}
			\ENDIF
			
			\STATE{updateWeights()}
            \FOR{dataset in datasets}
            	\STATE{ccLevel$_i$ = weight$_i$ $*$ numCh}
				\STATE{updateChannels()}
            \ENDFOR
        \ENDFOR
    \end{algorithmic}
\end{algorithm}

The energy-efficient target throughput algorithm aims at reaching the target throughput using as few channels as possible. It follows a simplified Finite State Machine with only 3 states, \textit{Slow Start}, \textit{Increase}, and \textit{Recovery}, in order to have a faster reaction time to changes in the channel.

After running through the \textit{Slow Start} phase (line 1-10), the algorithm measures the throughput and compares it with the target throughput. If it is higher or lower by a factory of $\beta$ or $\alpha$, respectively, it enters state \textit{Recovery}. After one more timeout, if the throughput is still higher than the target by at least a factor $\beta$, the channel count is reduced; on the other hand, if the throughput is lower than the target by at least a factor $\alpha$, the channel count is increased instead. No matter which feedback the algorithm received, it transitions back to state \textit{Increase} to keep measuring the achieved throughput. 

Finally, the algorithm updates the weights for each dataset and reassigns the channels across transfers similarly to the other algorithms.

\section{Experimental Results}
\label{sec:results}
\begin{table}[t]
    \centering
    \begin{tabular}{|m{1.3cm}|m{1.3cm}|m{0.7cm}|m{0.9cm}|m{2.3cm}|}
        \hline
        \textbf{Testbed} & \textbf{Bandwidth} & \textbf{RTT} & \textbf{BDP} & \textbf{CPU architecture} \\ \hline
        Chameleon Cloud & 10 Gbps & 32 ms & 40 MB & Haswell (server) \newline Haswell (client) \\ \hline
        CloudLab & 1 Gbps & 36 ms & 4.5 MB & Haswell (server) \newline Broadwell (client) \\ \hline
        DIDCLab & 1 Gbps & 44 ms & 5.5 MB & Haswell (server) \newline Bloomfield (client) \\ \hline
    \end{tabular}
    \caption{Characterics of testbeds}
    \label{tab:testbeds}
\end{table}

\begin{table}[t]
    \centering
    \begin{tabular}{|c|c|c|c|c|}
        \hline
        \textbf{Dataset} & \textbf{Num files} & \textbf{Total size} & \textbf{Avg file size} & \textbf{Std dev} \\ \hline
        Small files & 20,000 & 1.94 GB & 101.92 KB & 29.06 KB \\ \hline
        Medium files & 5,000 & 11.70 GB & 2.40 MB & 0.27 MB \\ \hline
        Large files & 128 & 27.85 GB & 222.78 MB & 15.19 MB \\ \hline
    \end{tabular}
    \caption{Characterics of datasets}
    \label{tab:datasets}
\end{table}

\begin{figure*}[ht]
    \centering
    \begin{subfigure}[t]{0.32\textwidth}
        \centering
        \caption{Average throughput (Chameleon Cloud)}
        \includegraphics[height=1.5in]{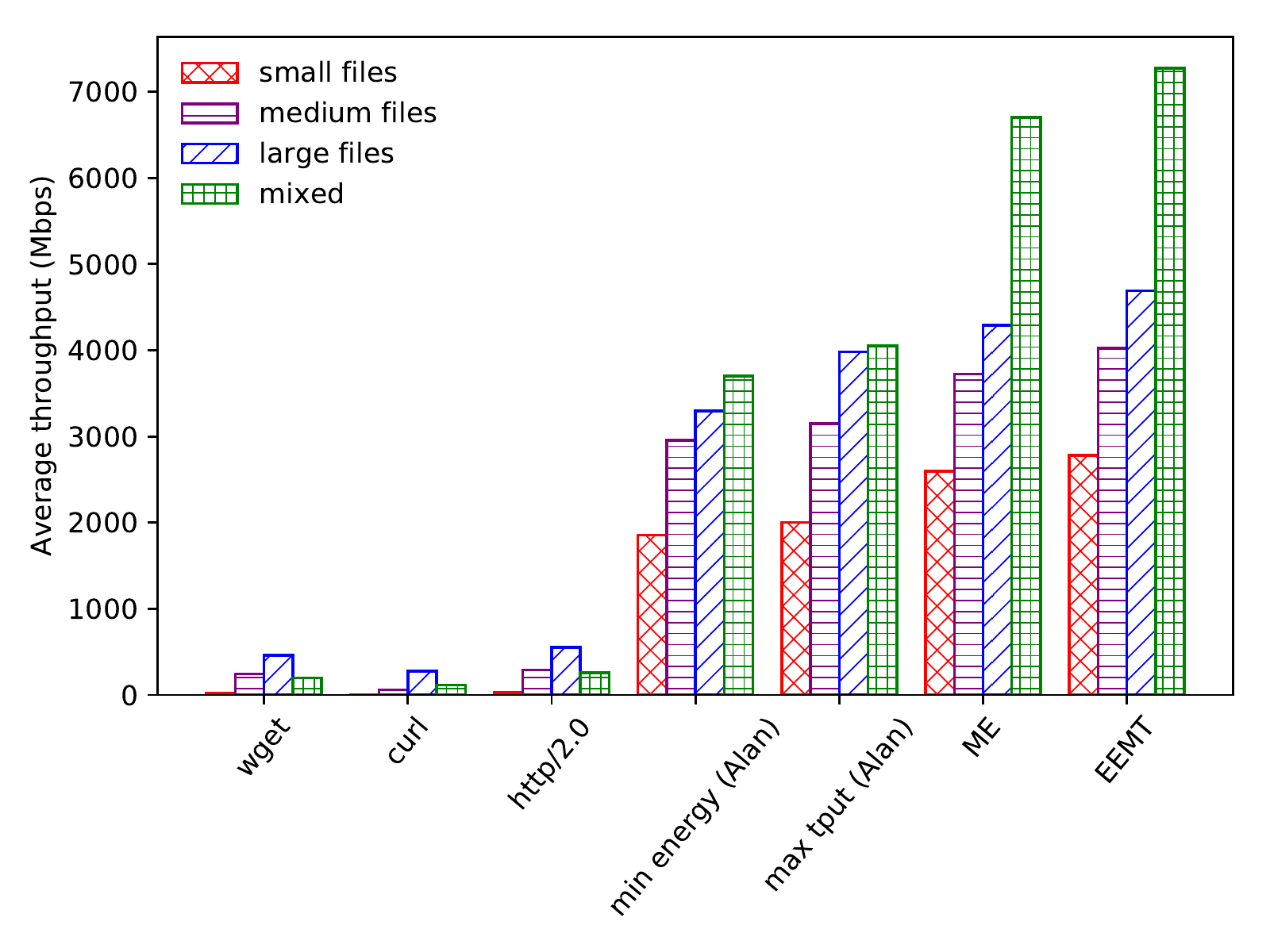}
    \end{subfigure}%
    ~ 
    \begin{subfigure}[t]{0.32\textwidth}
        \centering
        \caption{Average throughput (CloudLab)}
        \includegraphics[height=1.5in]{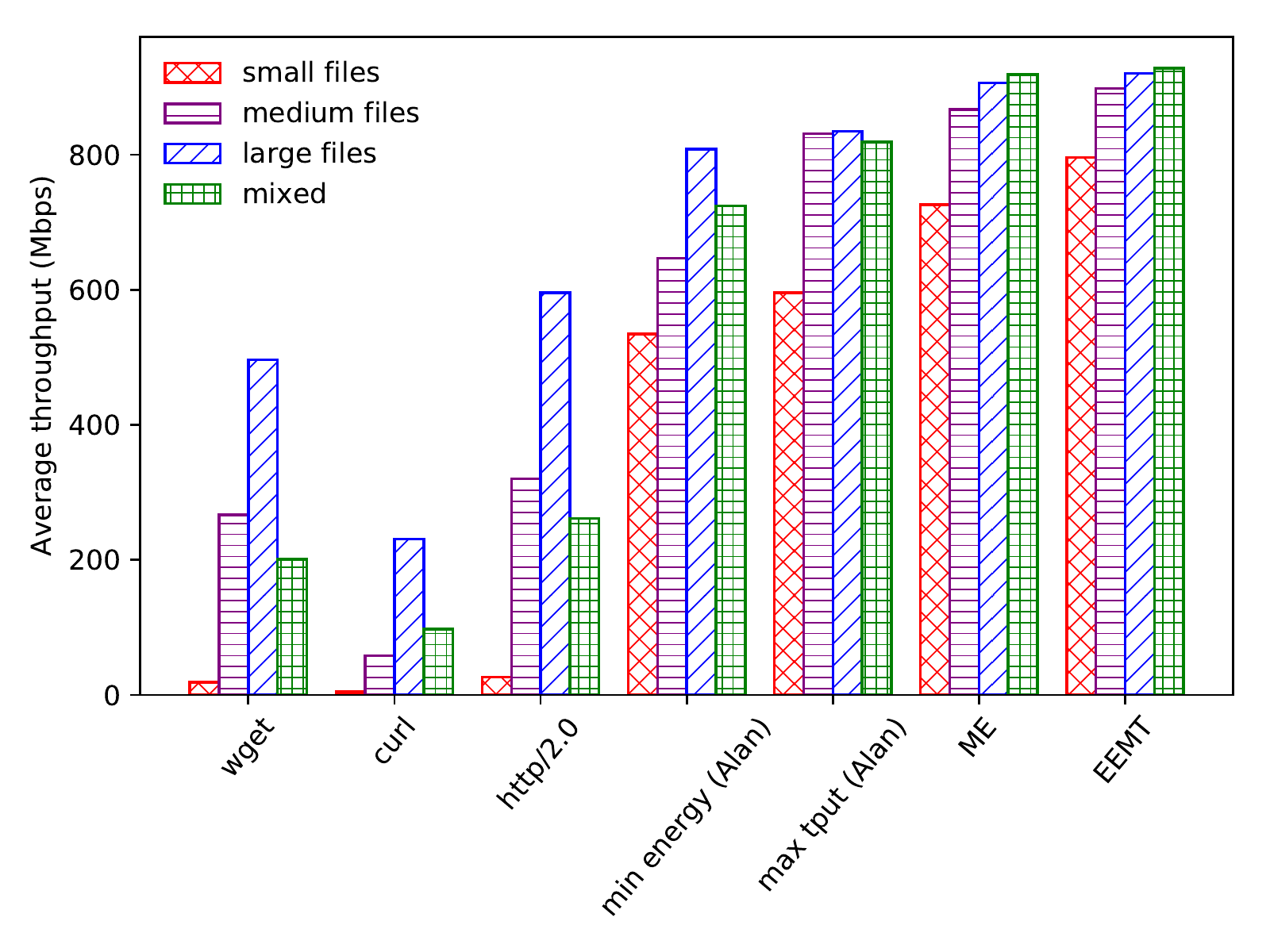}
    \end{subfigure}%
    ~ 
    \begin{subfigure}[t]{0.32\textwidth}
        \centering
        \caption{Average throughput (DIDCLAB)}
        \includegraphics[height=1.5in]{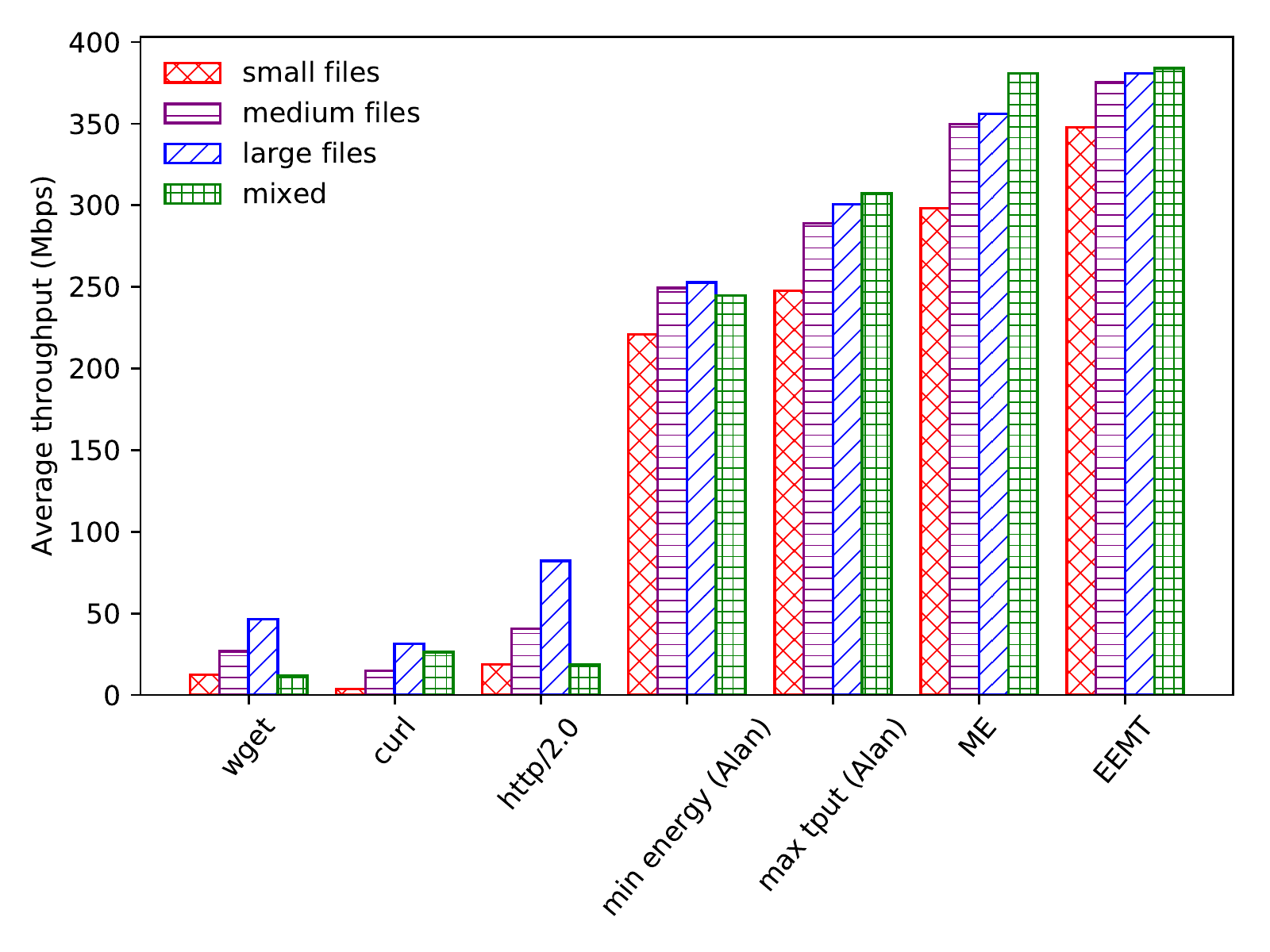}
    \end{subfigure}
    
    \begin{subfigure}[t]{0.32\textwidth}
        \centering
        \caption{Energy consumption (Chameleon Cloud)}
        \includegraphics[height=1.5in]{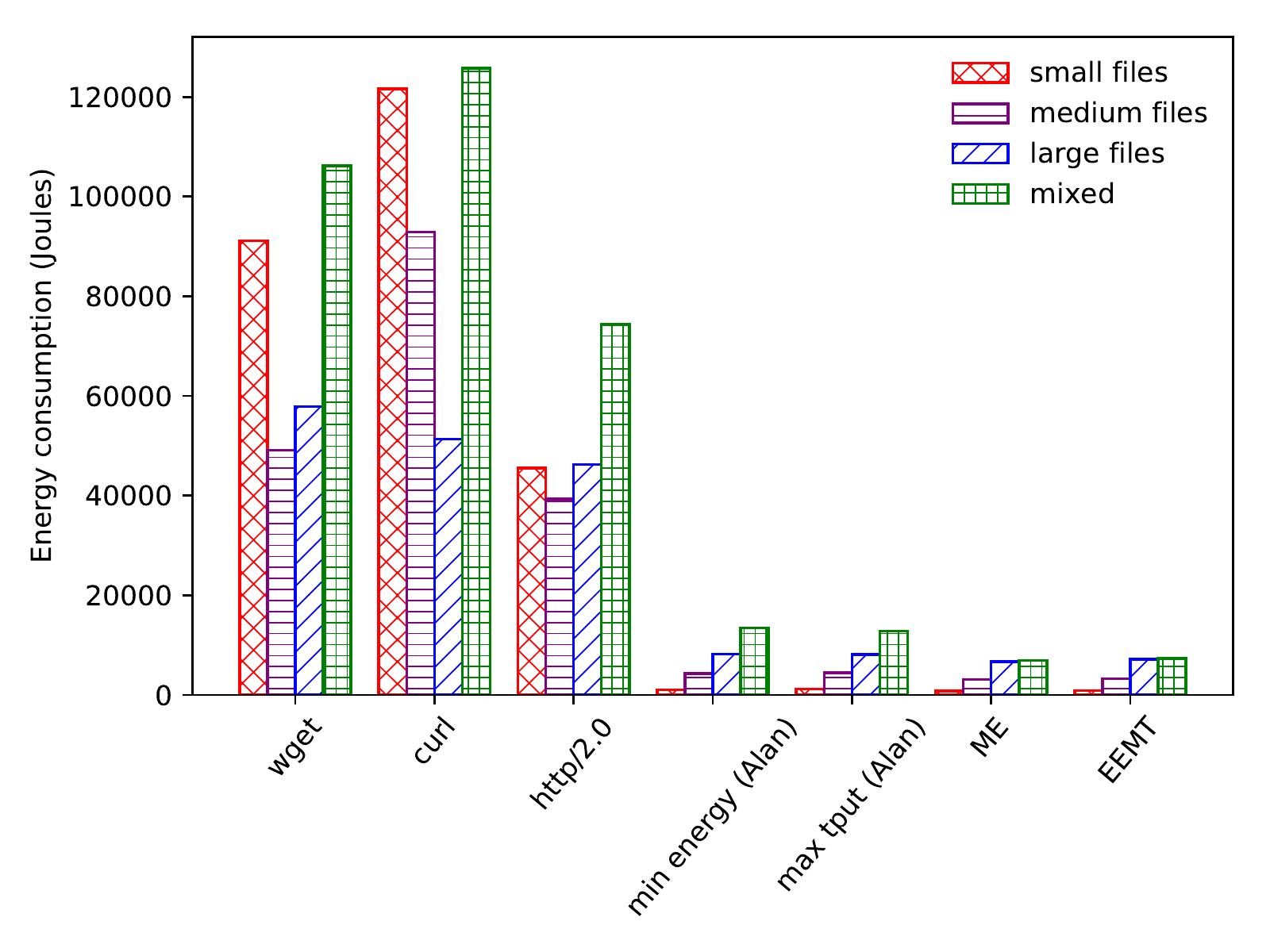}
    \end{subfigure}%
    ~ 
    \begin{subfigure}[t]{0.32\textwidth}
        \centering
        \caption{Energy consumption (CloudLab)}
        \includegraphics[height=1.5in]{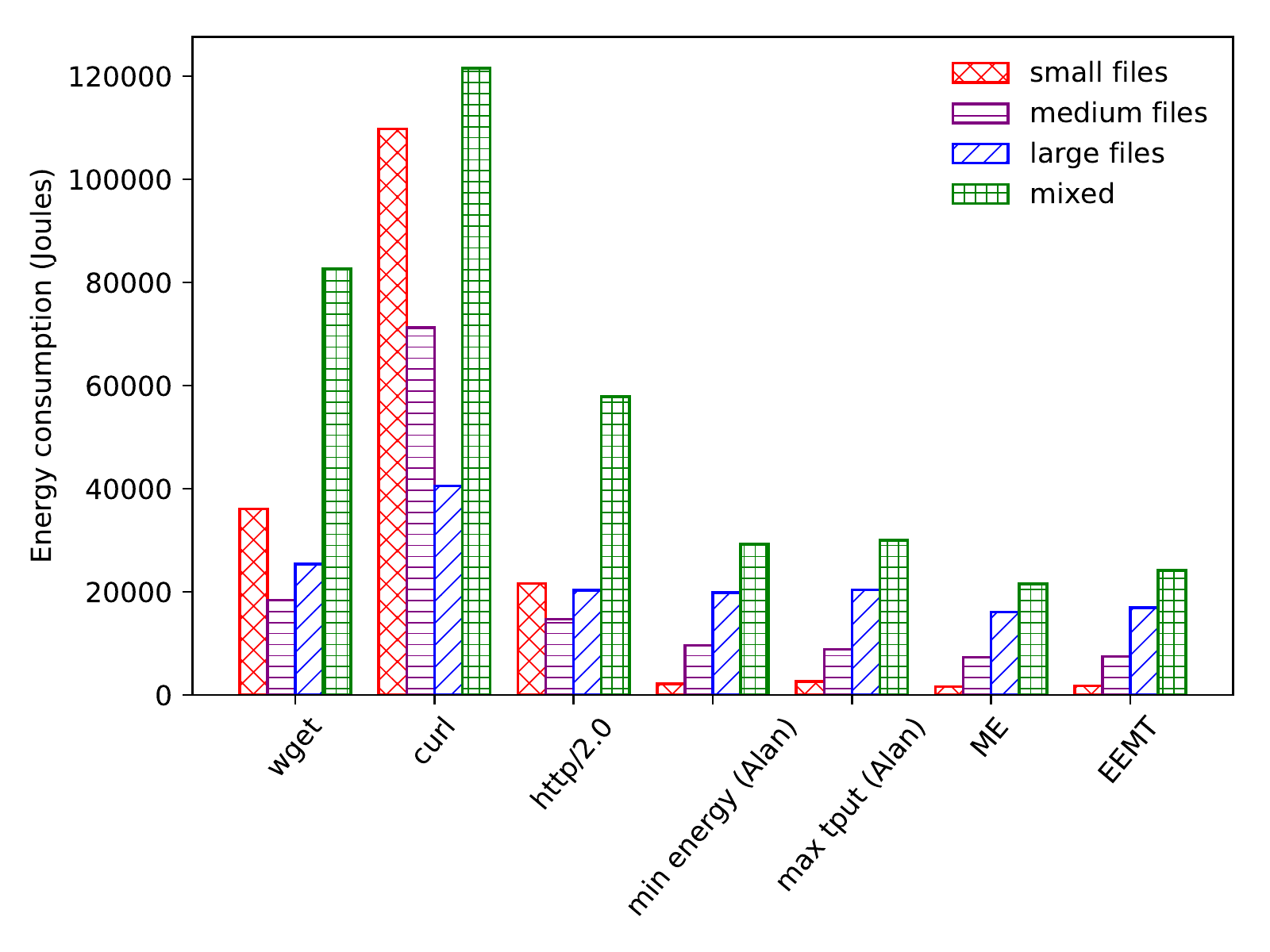}
    \end{subfigure}%
    ~ 
    \begin{subfigure}[t]{0.32\textwidth}
        \centering
        \caption{Energy consumption (DIDCLAB)}
        \includegraphics[height=1.5in]{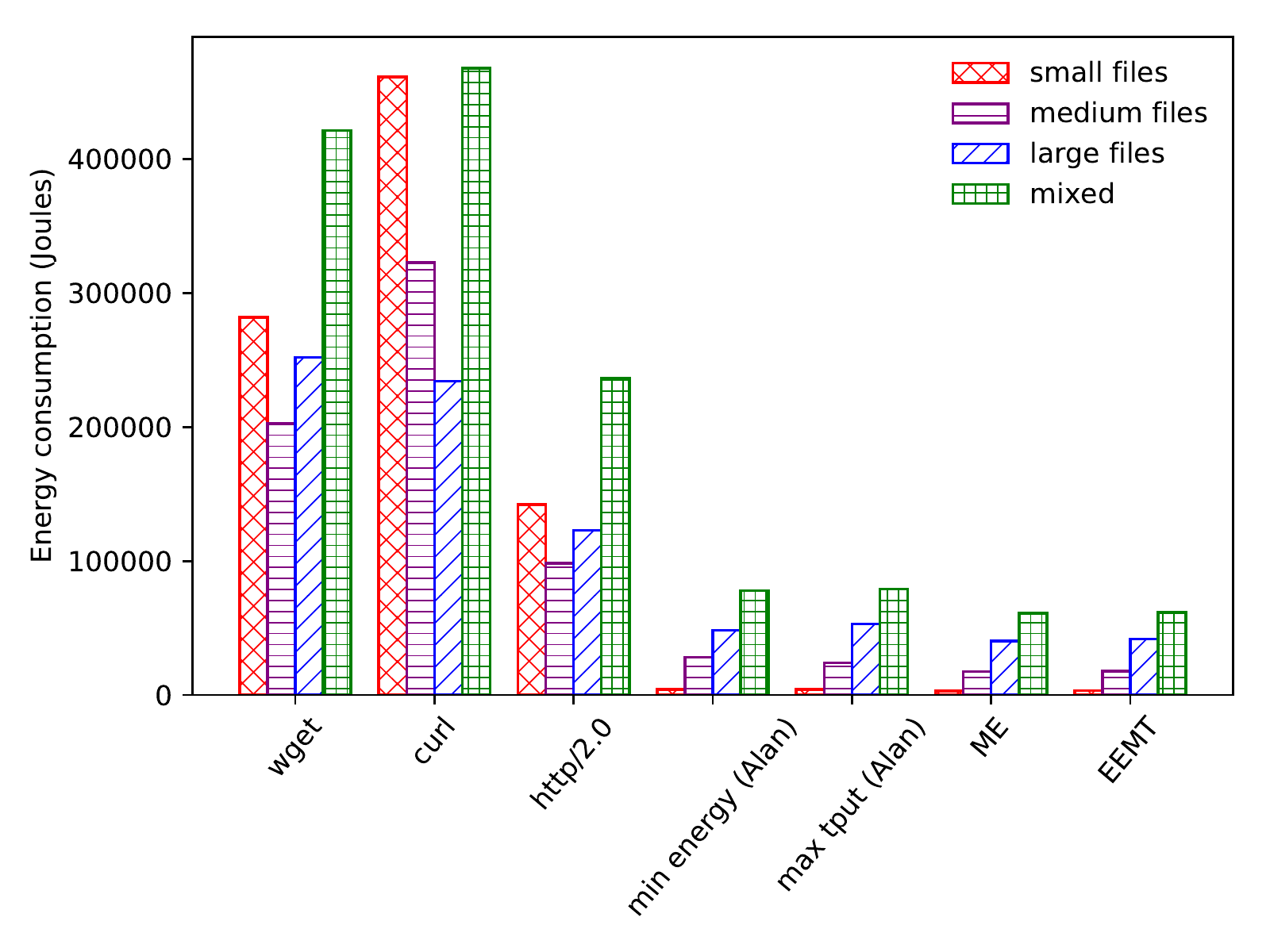}
    \end{subfigure}
    
    \caption{Comparison of throughput and energy consumption across 3 different testbeds}
    \label{fig:alg_comp}
\end{figure*}

We evaluated the parameter tuning algorithms on three different testbeds: i) Chameleon Cloud, with the server located at University at Chicago and the client at Texas Advance Computing Center; ii) CloudLab, with the server located at University of Wisconsin and the client at University of Utah; iii) DIDCLab, with the client located at the Data-Intensive and Distributed Computing Lab at University at Buffalo and the server at University at Chicago. An overview of the three testbeds is provided in table \ref{tab:testbeds}.

In order to compare the algorithms across different scenarios, we used four different datasets in the experiments: the three described in table \ref{tab:datasets}, and a mixed dataset, which is a combination of the previous three datasets.

We compared our algorithms with the only alternative solutions by Ismail et al. and some other commonly used data transfer tools: i) curl and wget, which are standard command line tools to transfer single files and datasets; ii) http/2.0, which is an upgrade to http/1.1 and promises to offer better performance by implementing multiplexing, which allows to transfer multiple streams over the same TCP connections.

The energy consumption has been measured using a Yokogawa WT210 digital power meter on the client in the DIDCLab testbed, while Intel RAPL has been used on every other node. Intel RAPL \cite{David2010} is a software power model that estimates energy usage by using hardware performance counters and I/O models. Its accuracy for the Haswell and Broadwell CPUs used in the experiments has been proved in previous work \cite{Khan2018,Hahnel2012,Zhang2014}. Since Intel RAPL provides accurate measurements for CPU and memory usage, we reduced the impact of disk activity to a minimum by performing all transfers memory-to-memory.


\begin{figure*}[ht]
    \centering
    \begin{subfigure}[t]{0.24\textwidth}
        \centering
        \caption{Average throughput with target (Chameleon Cloud)}
        \includegraphics[height=1in]{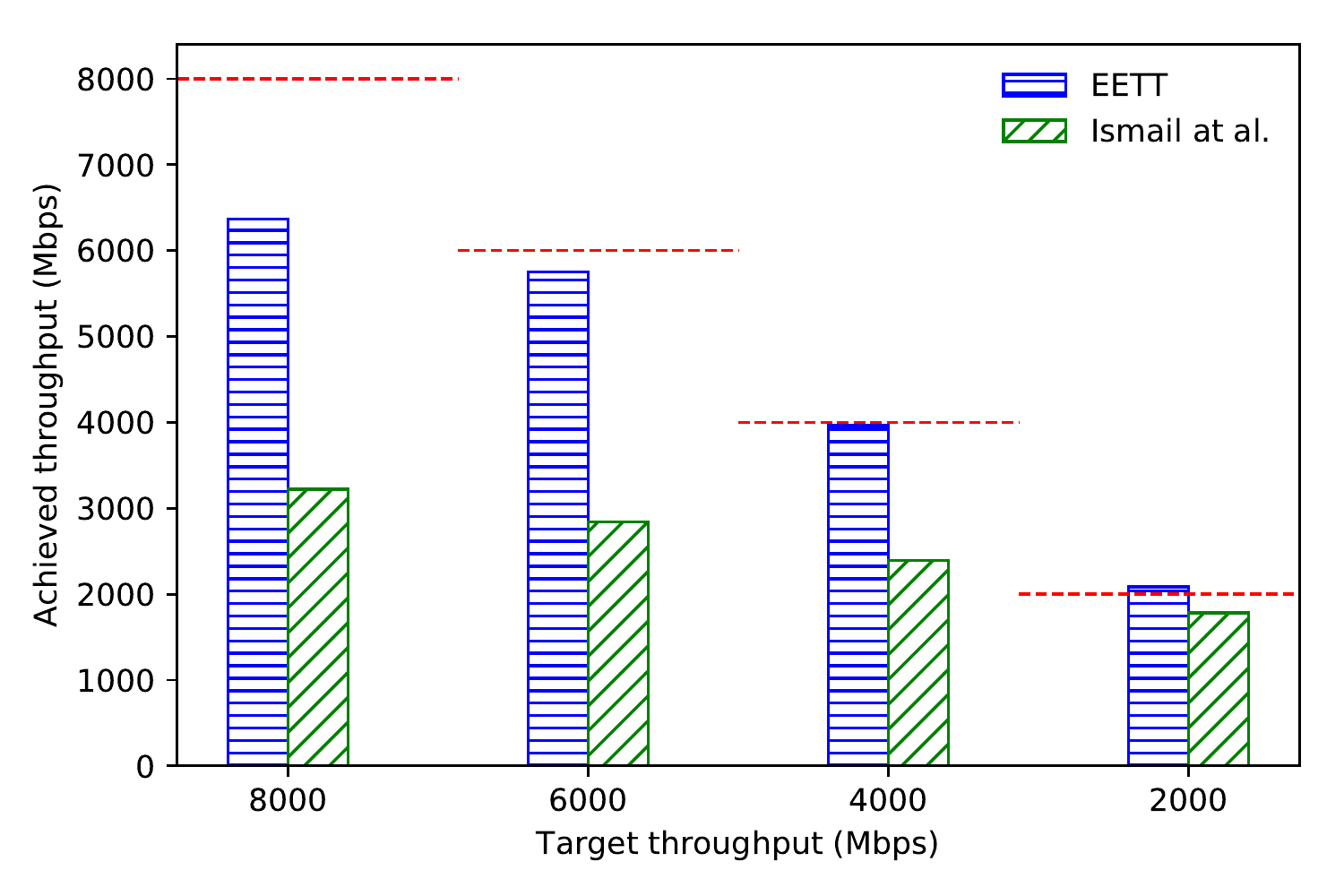}
    \end{subfigure}%
    ~ 
    \begin{subfigure}[t]{0.24\textwidth}
        \centering
        \caption{Average throughput with target (CloudLab)}
        \includegraphics[height=1in]{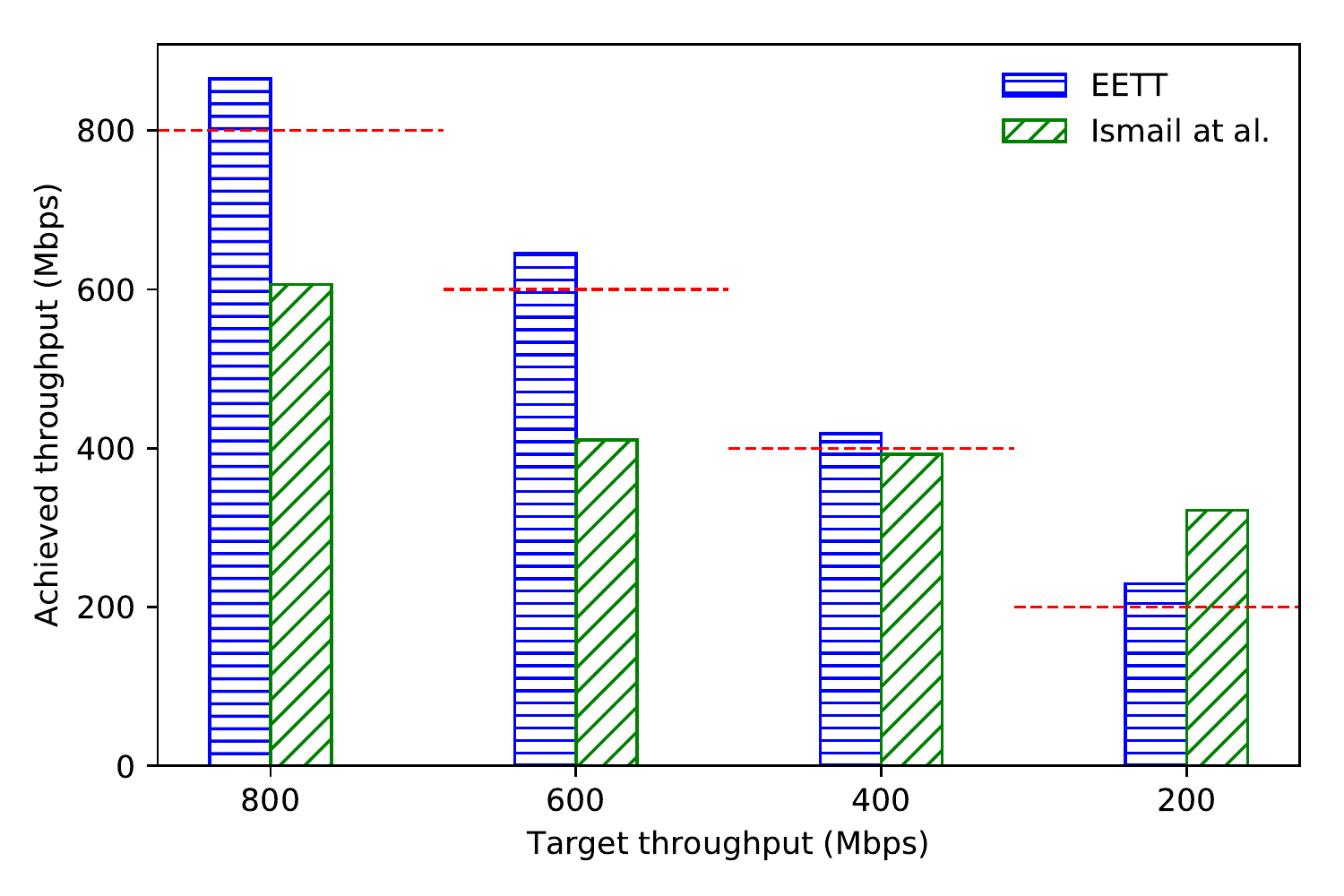}
    \end{subfigure}%
    ~
    \begin{subfigure}[t]{0.24\textwidth}
        \centering
        \caption{Energy consumption with target (Chameleon Cloud)}
        \includegraphics[height=1in]{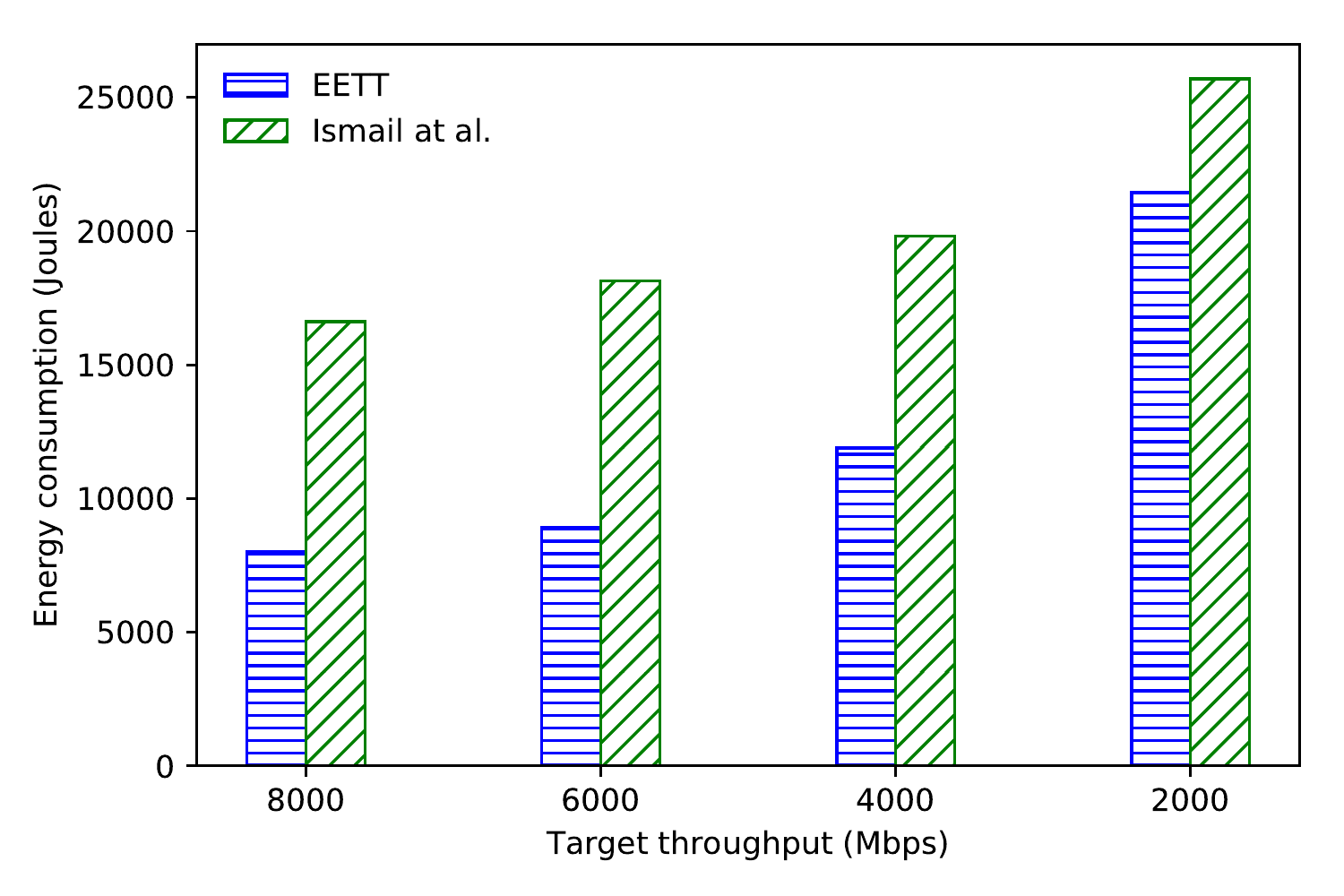}
    \end{subfigure}%
    ~ 
    \begin{subfigure}[t]{0.24\textwidth}
        \centering
        \caption{Energy consumption  with target (CloudLab)}
        \includegraphics[height=1in]{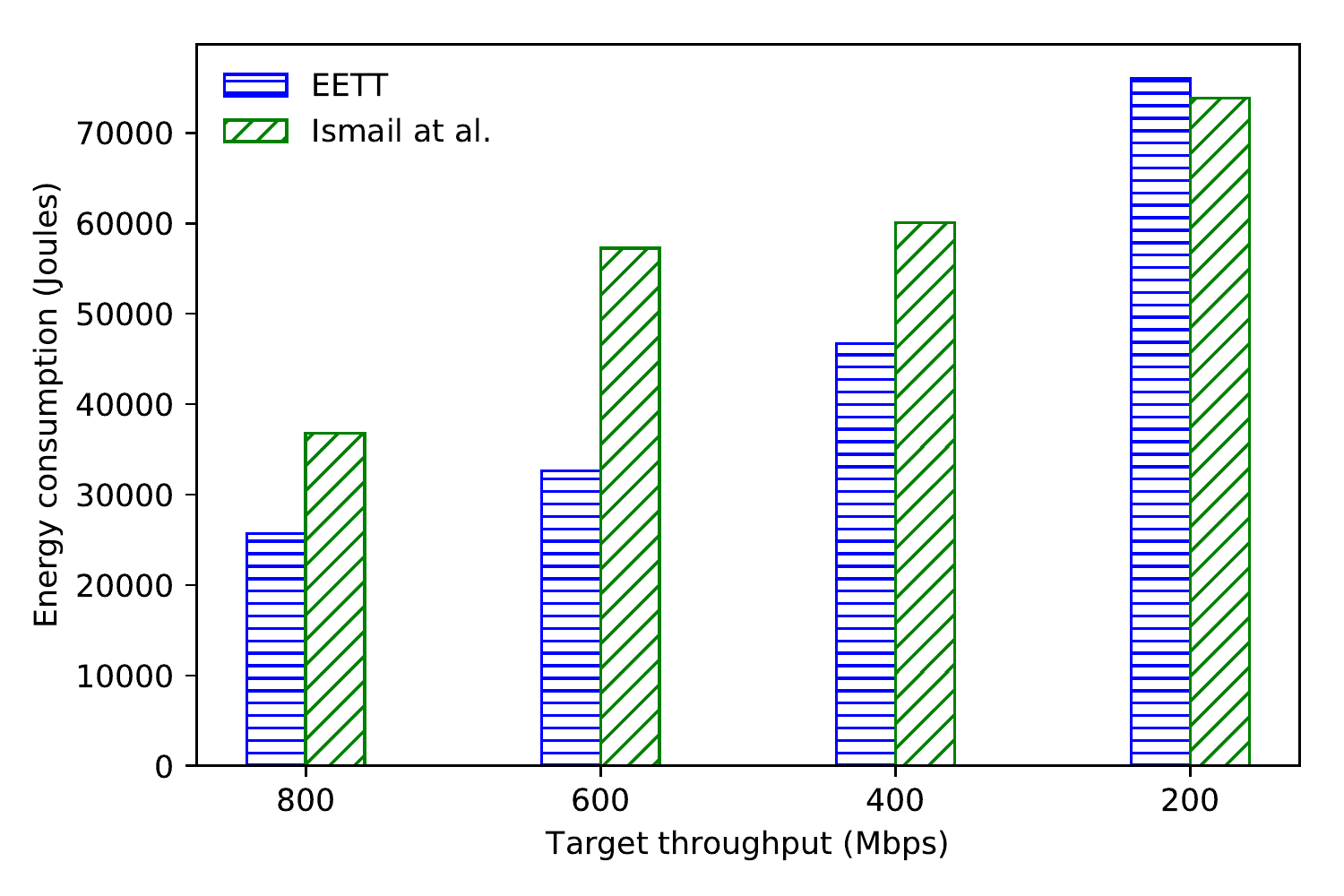}
    \end{subfigure}
    
    \caption{Comparison of target throughput algorithms}
    \label{fig:target_comp}
    \vspace{-3mm}
\end{figure*}

\subsection{Comparison of minimum energy and maximum throughput algorithms}

Figure \ref{fig:alg_comp} shows the throughput and energy consumption of the transfer tools and algorithms tested on the Chameleon Cloud, CloudLab, and DIDCLAB testbeds.

Wget and curl perform very poorly due to the lack of any optimization, with high energy consumption and very low throughput. On the other hand, http/2.0 achieves better performance thanks to multiplexing, which reduces the impact of RTTs, especially when transferring small files. However, on a wide area network, http/2.0 is not able to fully use the bandwidth due to the lack of parallelism and concurrency tuning.

Conversely, Minimum Energy and Maximum Throughput algorithms by Ismail et al. achieve much better performance, although they suffer from some major drawbacks, which are evident when transferring the large and mixed datasets on the large BDP testbed (Chameleon Cloud): i) static parameter tuning, which at times leads to suboptimal parameters; ii) in both algorithms, as the buffer size grows to match the network BDP, the parallelism level drops to 1, causing poor performance.

On the other hand, our algorithms outperform the state-of-the-art solutions across all different scenarios. ME reduces the energy consumption by up 48\% with respect to Min Energy algorithm by Ismail et al. when transferring the mixed dataset. Moreover, EEMT achieves better throughput by up to 80\% when transferring the mixed dataset compared to Maximum Throughput algorithm by Ismail et al. and reduces the energy consumption by up to 43\%.

The reduced energy consumption is mainly due to better dynamic tuning of the transfer parameters and the use of dynamic scaling of CPU frequency and number of active cores.

\subsection{Comparison of target throughput algorithms}

Figure \ref{fig:target_comp} shows a comparison between our target throughput algorithm and the state-of-the-art solution by Ismail et al., using the mixed datasets and different target values (80\%, 60\%, 40\%, and 20\% of the maximum theoretical bandwidth). We excluded the DIDCLab testbed from the comparison, due to the low available bandwidth.

Our target throughput algorithm achieves a throughput withing 5-10\% of the target across all scenarios, with the only exception of Chameleon Cloud when the target is set to 8 Gbps. However, since no algorithm achieves more than 7 Gbps (figure \ref{fig:alg_comp}, it is most likely due to low available bandwidth. The target throughput algorithm by Ismail et al. is able to achieve the target for low throughput values (20\% of the bandwidth on Chameleon Cloud and 20-40\% on CloudLab). The poor performance is most likely due to 2 factors: i) the algorithm starts with one channel and slowly increments its channel count, taking a very long time to achieve the target; ii) the algorithm does not distribute the channels across datasets based on the remaining size or current speed, resulting in slower datasets becoming bottlenecks.

Even when achieving a very similar throughput, our algorithm consumes much less energy than the target throughput algorithm by Ismail et al., achieving 20\% reduced energy consumption on Chameleon Cloud for a target of 2 Gbps and 29\% less energy consumption on CloudLab for a target of 400 Mbps.

The only scenario in which the algorithm by Ismail et al. consumes slightly less energy than EETT is on CloudLab when the target is set to 200 Mbps; however, this is due to the fact that it achieves a throughput 60\% higher than the target, greatly reducing the transfer time. 

\begin{figure*}[ht]
    \centering
    \begin{subfigure}[t]{0.32\textwidth}
        \centering
        \caption{Energy consumption (Chameleon Cloud)}
        \includegraphics[height=1.5in]{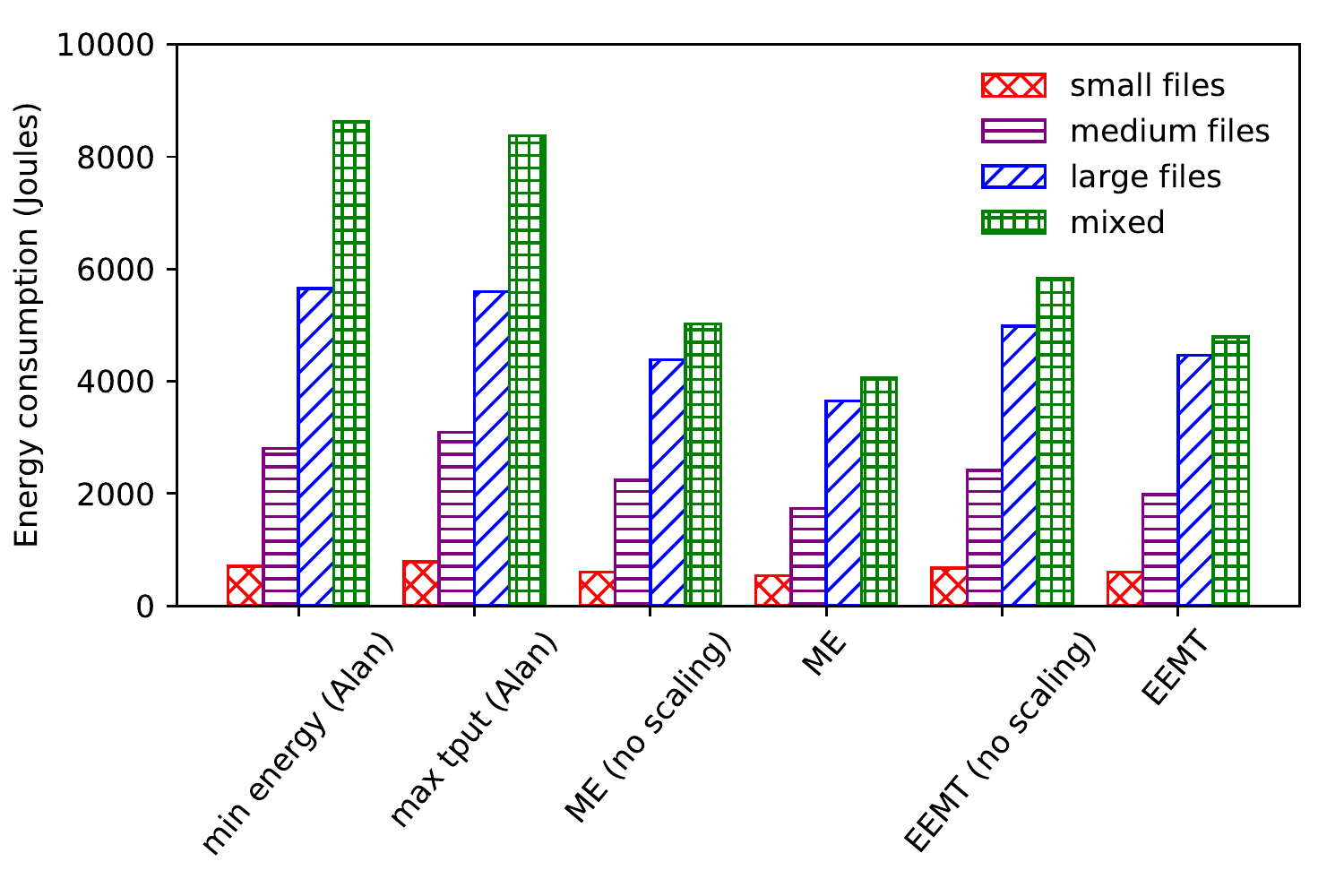}
    \end{subfigure}%
    ~ 
    \begin{subfigure}[t]{0.32\textwidth}
        \centering
        \caption{Energy consumption (CloudLab)}
        \includegraphics[height=1.5in]{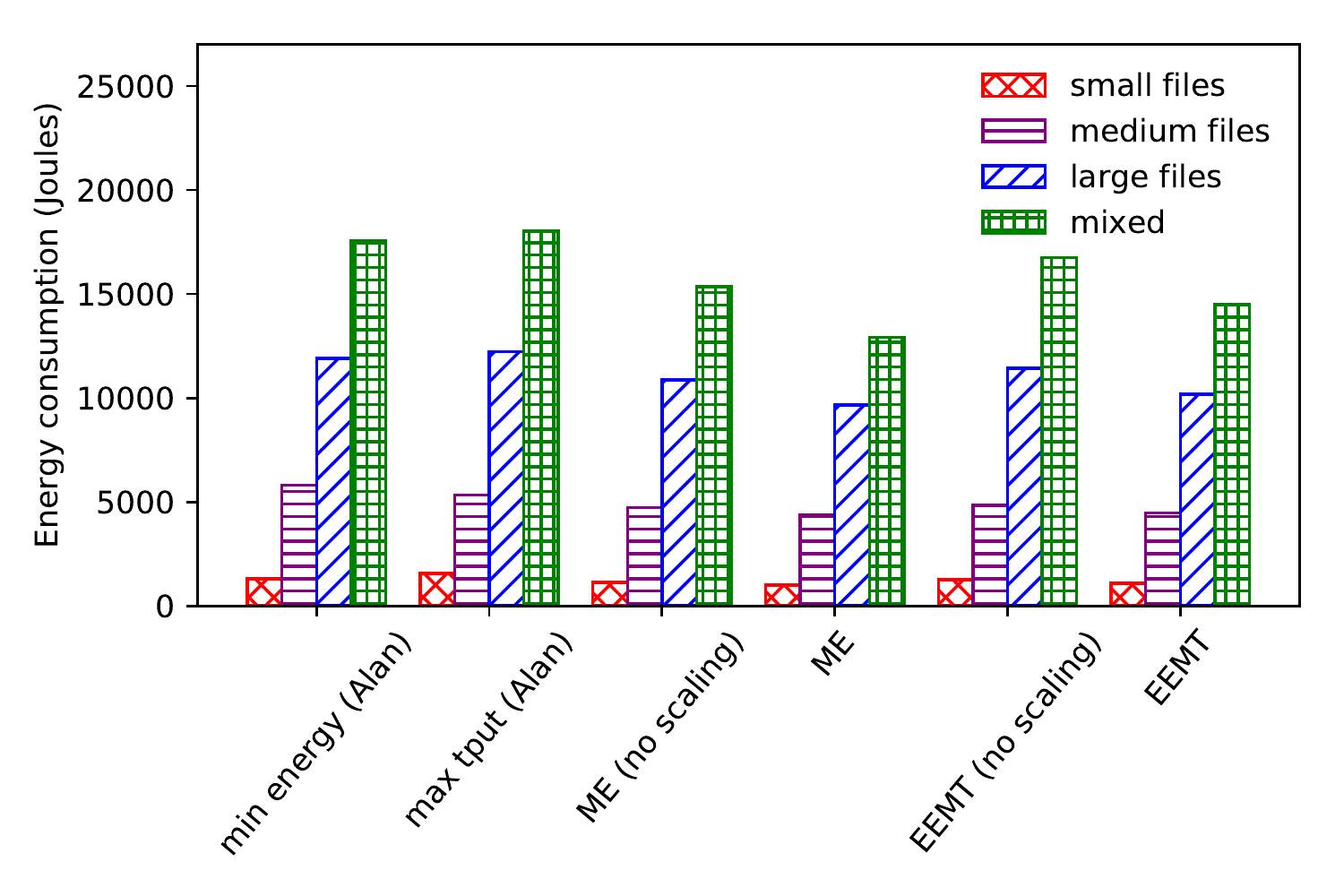}
    \end{subfigure}%
    ~ 
    \begin{subfigure}[t]{0.32\textwidth}
        \centering
        \caption{Energy consumption (DIDCLAB)}
        \includegraphics[height=1.5in]{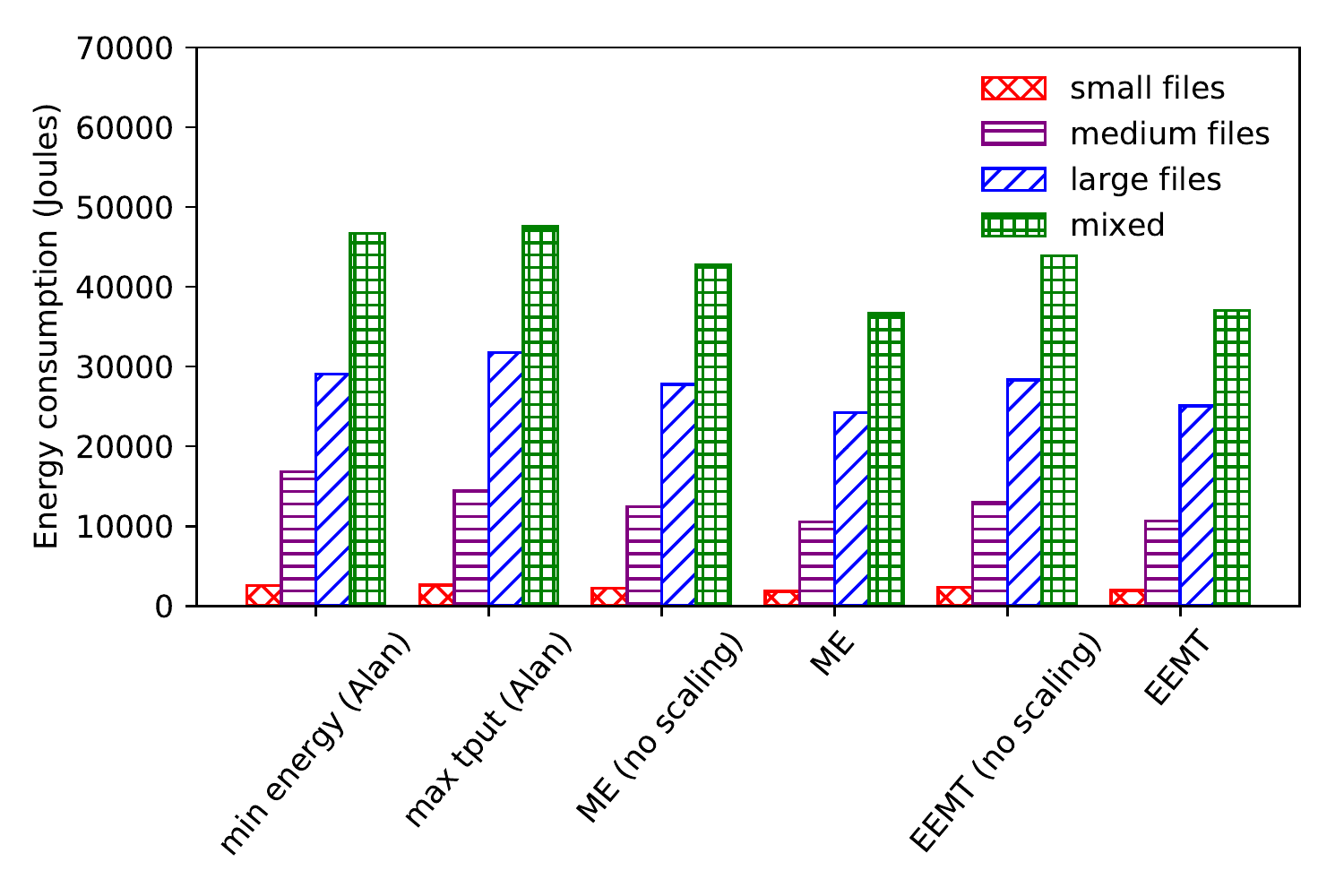}
    \end{subfigure}
    
    \caption{Effect of frequency and core scaling on the client's energy consumption}
    \label{fig:noscaling_comp}
\vspace{-3mm}
\end{figure*}

\subsection{Effect of frequency scaling and number of active cores}

In order to analyze the extent to which CPU frequency and core scaling reduce the energy consumption, we removed the load control module from the 2 algorithms ME (Minimum Energy) and EEMT (Energy Efficient Maximum Throughput). We then measured the energy consumption on the client, since there is no frequency scaling on the server.

Figure \ref{fig:noscaling_comp} shows a comparison of all algorithms across the 3 testbeds. Without frequency and active core scaling, ME reduces the energy consumption by up to 42\% on Chameleon Cloud with respect to Min Energy algorithm by Alan et al., while EEMT achieves 30\% less energy consumption than Max Throughput algorithm by Alan et al on the same testbed. However, when adding frequency and core scaling to both algorithms, the energy consumption drops by an additional 19\% on ME and 17\% on EEMT, bringing the total energy reduction to 53\% with respect to Min Energy (Alan et al.) and 43\% with respect to Max Throughput (Alan et al.).

On the DIDCLab testbed, the bandwidth is much more limited, and therefore the potential for energy saving is lower than on a large bandwidth testbed such as Chameleon Cloud. In fact, ME's energy consumption is 9\% lower than Min Energy by Alan et al. with no scaling. However, when using frequency and active core scaling, the energy saving rises to 22\%, showing the potential of such technique on limited bandwidth scenarios. Similarly, EEMT achieves 8\% less energy consumption than Max Throughput by Ismail et al. without scaling, whereas the reduction in energy rises to 23\% when using CPU frequency and core scaling.

\section{Related Work}
\label{sec:related_work}
Application layer network optimization mainly focuses on tuning protocol parameters to avoid congestion. 
%
%
%
Earlier works proposed models that allocate TCP socket buffer to saturate the network links~\cite{jain2003tcp}.
%
%
However, TCP buffer size allocation alone fails to achieve the optimal bandwidth in a long RTT networks. Numerous work has been proposed to open multiple parallel streams to increase the transfer throughput~\cite{lu2005modeling,Yildirim:2009:BTB:1552280.1552283, JGrid_2012, Thesis_2005, CCPE_2006}. 
%
%
Liu et al.~\cite{R_Liu10} introduced a GridFTP based solution that can open multiple transfer sessions to facilitate concurrent the file transfers from a single dataset.
Pipelining~\cite{TCP_Pipeline} was introduced to reduce the one RTT delay between each small file transfers. 
%
%
%

%
Energy efficiency in data communication deals mostly with network infrastructure. Many works suggested different power modes~\cite{Gupta_2003}. 
%
%
%
S. Nedevshi et al.~\cite{Nedevschi_2008} explored the joint effect of sleeping support and network rate adaptation based on workloads.
Hardware level energy efficiency was proposed at 802.3az standards ~\cite{IEEE_802} to make ethernet cards more energy efficient. 
%
%
%
Alan et al. investigated the energy consumption and throughput of data transfer under different concurrency and parallelism levels. They proposed a heuristic based parameter search to improve performance and energy consumption~\cite{Alan2015, Kosar_jrnl14}.

\section{Conclusion}
\label{sec:conclusion}
In this paper, we introduced three novel application-level parameter tuning algorithms to provide SLA-based energy-efficient data transfer service. The algorithms combine heuristics and runtime tuning to satisfy the SLA requirements set by the user. Our model reduces the energy consumption by dynamically tuning the CPU frequency and changing the number of active cores, as well as adjusting the pipelining, parallelism, and concurrency levels. Experimental results show that the proposed algorithms outperform the state-of-the-art solutions, providing up to 48\% reduced energy consumption and 80\% better throughput.

\bibliographystyle{abbrv}
\bibliography{main,didc,misc,new_refs} 

\end{document}